\documentclass[prl,aps,floats,twocolumn,showpacs,
preprintnumbers,showkeys]{revtex4-1}

\usepackage[latin1]{inputenc}                    
\usepackage{graphicx}                            
\usepackage{latexsym}                            
\usepackage{amsfonts}                            
\usepackage{amssymb}                             
\usepackage{amsmath}                             
\usepackage[mathscr]{eucal}                      
\usepackage{dcolumn}                             
\usepackage{multirow}                            
\usepackage{theorem}                             
\usepackage[normalem]{ulem}                      
\usepackage{psfrag}

\newcommand{\bc}{\begin{center}}
\newcommand{\ec}{\end{center}}
\newcommand{\be}{\begin{equation}}
\newcommand{\ee}{\end{equation}}
\newcommand{\bea}{\begin{eqnarray}}
\newcommand{\eea}{\end{eqnarray}}

\newcommand{\emdash}{\hspace{1pt}---\hspace{1pt}}

\begin{document}

\preprint{
\vbox{
\hbox{ADP-12-12/T779}
}}

\title[Neutron Star Properties with Hyperons]{Neutron Star Properties with Hyperons}
\author{D.~L.~Whittenbury}
\author{J.~D.~Carroll}
\author{A.~W.~Thomas}
\author{K.~Tsushima}
%
\affiliation{CSSM and ARC Centre of Excellence for Particle Physics at the 
Terascale,\\ School of Chemistry and Physics,
University of Adelaide,
  Adelaide SA 5005, Australia}
\author{J.~R.~Stone} 
\affiliation{Department of Physics, University of
  Oxford, Oxford OX13PU, United Kingdom}

\begin{abstract}
In the light of the recent discovery of a neutron star with a mass
accurately determined to be almost two solar masses, we carefully 
examine the maximum mass of a neutron star containing hyperons within 
a relativistic Hartree-Fock treatment. Using the quark-meson coupling 
model, which naturally incorporates hyperons without additional 
parameters, we find a maximum mass a little over 2.1 $M_{\odot}$. 
\end{abstract}



\maketitle
\raggedbottom
%
The recent observation of a $1.97\pm 0.04 \, M_{\odot}$ millisecond
pulsar, PSR J1614-2230, by Demorest \textit{et
  al.}~\cite{Demorest:2010bx} has set the most stringent limit on
models of neutron star cores so far. This discovery has spurred a
re-examination of the possibility of exotica such as hyperons, Bose
condensates, and quark matter playing an important role in models of
neutron star interiors, owing to a presumed softening of the equation
of state (EoS) expected in the presence of additional degrees of
freedom.  Historically, this has led to expectations of reduced
maximum neutron star masses for compact objects in hydrostatic
equilibrium.

In this Letter we build on the earlier work of 
Stone {\it et al.}~\cite{RikovskaStone:2006ta}, who already 
predicted the existence of neutron stars containing hyperons with 
masses as large as 2 $M_{\odot}$ in 2007, to establish a conservative 
upper limit on the maximum mass of such a star.  
We work within the quark-meson coupling (QMC)
model~\cite{Guichon:1987jp,Guichon:1995ue,Guichon:2008zz}, 
which has the advantage of being derived from the quark level, 
with a very small number of adjustable parameters, while being 
consistent with a broad range of constraints derived from hypernuclei 
as well as normal nuclear properties. We find that the stability 
under variation of the very small number of adjustable parameters
is such that if a star were discovered with a mass significantly above
$2.1~M_{\odot}$, we would need to consider more exotic physics, 
because it could not be accommodated within the QMC model.

We recall that 
QMC is based upon the self-consistent modification of the
structure of a baryon embedded in nuclear matter.
At Hartree level it involves only three adjustable parameters
which describe the effective couplings of the $\sigma$, $\omega$ 
and $\rho$ mesons to the $u$ and $d$ quarks. These are fixed 
by adjusting them to fit the properties of
symmetric nuclear matter, namely its saturation density and 
binding energy as well as its symmetry energy.
We note that the $\sigma$
meson used here simply serves as a convenient representation of the
scalar-isoscalar attraction arising from two-pion exchange.  

In the most recent development of the QMC model~\cite{Guichon:2008zz},
the self-consistent inclusion of the gluonic hyperfine interaction led
to a very successful description of the binding energies of
$\Lambda$-hypernuclei\emdash as well as the observed absence of medium
and heavy mass $\Sigma$-hypernuclei\emdash with no additional
parameters. We stress that this is achieved without any coupling of 
the strange quark to the $\sigma, \omega$ and $\rho$ mesons 
(which would be OZI suppressed) and without the need to introduce any 
further mesons. While the model could be supplemented with much 
heavier mesons containing strange quarks~\cite{Weissenborn:2011ut}, 
Occam's razor suggests 
that one should not introduce them if they are not needed.

A clear connection has been established between the self-consistent
treatment of in-medium hadron structure and the existence of
many-body~\cite{Guichon:2004xg} or density
dependent~\cite{Guichon:2006er} effective forces.  Dutra {\it et
  al.}~\cite{Dutra:2012mb} critically examined a variety of
phenomenological Skyrme models of the effective density dependent
nuclear force against the most up-to-date empirical constraints. 
Amongst the few percent of the Skyrme forces studied which satisfied 
all of these constraints, the Skyrme model SQMC700, derived
from the QMC model, was unique in that it incorporated the effects of
the internal structure of the nucleon and its modification in-medium.

While the earlier study of Stone {\it et al.}~\cite{RikovskaStone:2006ta} 
demonstrated the
importance of exchange (Fock) terms in calculations of the EoS of
dense baryonic matter in $\beta$-equilibrium, it included only the
Dirac vector term in the vector-meson-nucleon vertices.  
In this work we include the full vertex
structure which one might expect to enhance the pressure at high
density.  This is especially so in the case of the $\rho$ meson for
which the tensor coupling is much larger than that of the $\omega$.
Our calculation extends the work of Krein {\it et
  al.}~\cite{Krein:1998vc}, who considered nucleons only, by
evaluating the full exchange terms for all octet baryons and adding
them in the same way as Stone {\it et
  al.}~\cite{RikovskaStone:2006ta}; as additional contributions to the
energy density.  
It also extends and complements the important work of Miyatsu {\it et al.} 
who calculated the properties of neutron stars within an earlier version 
of the QMC model~\cite{Miyatsu:2011bc}. In particular, we employ the latest 
version of the model which reproduces key hypernuclear properties without 
the adjustment of coupling constants needed there. Furthermore, 
we carefully explore 
the limit on the maximum mass of a neutron star containing hyperons 
while ensuring consistency with critical nuclear properties, 
such as the incompressibility 
of nuclear matter.

Next we describe the main features of the QMC model and
explore its parameters arranged into groups (``scenarios'') to test
the robustness of its predictions. Within the QMC model, the
baryon energy density, $\epsilon_B$, is given by
\vspace{-2mm}
\begin{equation}
\epsilon_{B} = \frac{2}{(2\pi)^{3}}\sum_{B}
\int\limits_{\vert \mathbf{p}\vert <p_{F}}\!\!\!\!\! d\mathbf{p}\; 
\sqrt{p^{2}+M^{\ast\, 2}_{B}} \, , \  
\vspace{-2mm}
\end{equation}
where the effective, in-medium baryon masses, $M^{\ast}_{B}$, are
calculated self-consistently for an MIT bag immersed in (and in
Ref.~\cite{Guichon:2008zz}, parameterized as functions of) a mean
scalar field, designated here by a barred symbol. At a given density,
$\bar{\sigma}$ is self-consistently expressed as
\vspace{-2mm}
\be 
\bar{\sigma} 
= - \frac{2}{m^{2}_{\sigma}(2\pi)^{3}}
\sum_{B}\int\limits_{\vert \mathbf{p}\vert <p_{F}}\!\!\!\!\!
d\mathbf{p}\; 
\frac{M^{\ast}_{B}}{\sqrt{p^{2}+M^{\ast\, 2}_{B}}}
\frac{\partial
    M^{\ast}_{B}}{\partial\bar{\sigma}}\ \, . 
    \vspace{-2mm}
\ee 
An additional contribution, $\delta \bar{\sigma}$, to the scalar field
arises if we include the Fock terms in the minimization 
of the energy density. This
provides only a small correction to the mean field, and its effect is
included only in the scenario denoted $\delta \bar{\sigma}$.  The total
hadronic energy density, $\epsilon_{H}$, is the sum of baryonic,
$\epsilon_{B}$, and mesonic, $\epsilon_{\sigma \omega \rho \pi}$
contributions, for which
%
\begin{eqnarray} \label{eq:energy}
\epsilon_{\sigma \omega \rho \pi} 
&=& \sum_{\alpha=\sigma,\omega,\rho}\frac{1}{2}m^{2}_{\alpha}
\bar{\alpha}^{2} \nonumber \\
&&+\sum_{\alpha=\sigma,\omega,\rho,\pi}\ \sum_{BB'} 
\frac{C^{\alpha}_{BB'} }{(2\pi)^{6}}
\iint\limits_{\substack{\vert \mathbf{p}\vert <p_{F}\\
| \mathbf{p}'| <p_{F'}} }\!\!\!\! d\mathbf{p}\; d\mathbf{p}'\, 
\mathbf{\Xi}_{BB'}^{\alpha}\, ,
\end{eqnarray}
where $C^{\sigma}_{BB'}=C^{\omega}_{BB'}=\delta_{BB'}$.
$C^{\rho}_{BB'}$ and $C^{\pi}_{BB'}$, which arise from symmetry
considerations, are given in Ref.~\cite{RikovskaStone:2006ta}.  Note
that the $\pi$ meson only contributes to the second term in
Eq.~(\ref{eq:energy}) as it is coupled via a pseudo-vector current. For
$\epsilon_{\sigma\omega\rho}$, the integrand has the form
\vspace{-2mm}
\be
\mathbf{\Xi}^{\alpha}_{BB'}  = \frac{1}{2}\sum_{s,s'} \vert
\bar{u}_{B'}(p',s')\Gamma_{\alpha} u_{B}(p,s)\vert^{2} 
\Delta_{\alpha}(\mathbf{k})\, ,
\vspace{-2mm}
\ee
where $\Delta_{\alpha}(\mathbf{k})$ is the Yukawa propagator for meson
$\alpha$ with momentum $\mathbf{k} =\mathbf{p} - \mathbf{p'}$. For the
vector mesons, the full vertex structure is included in the manner of
Ref.~\cite{Krein:1998vc} as
\vspace{-2mm}
\be 
\Gamma_{\alpha B} = \epsilon^{\mu}_{\alpha}\left[
g_{\alpha B} \gamma_{\mu} F^{\alpha}_{1}(k^{2}) 
+\frac{i f_{\alpha B}\sigma_{\mu\nu}}{2M^{\ast}_{B}}k^{\nu} 
F^{\alpha}_{2}(k^{2})\right] \, . 
\vspace{-2mm}
\ee 

As usual, the effect of short distance repulsion on the Fock terms is
simulated by the replacement
\be
\frac{\vec{q}^{\; 2}}{(\vec{q}^{\; 2} + m^2)}\ \to \ 1  -
\frac{m^2}{(\vec{q}^{\; 2}+m^2)}
\ee
from which the unit term is subtracted, thus eliminating a
$\delta$-function.
The form factors $F_{1,2}^{\alpha}$ all have the same dipole form with
the cutoff mass $\Lambda$ varied from 0.9 to 1.3 GeV to test the
sensitivity. As a further test of the model dependence, we consider
two choices for the ratios of tensor to vector coupling constants
${\displaystyle \kappa_{\alpha B} = f_{\alpha B}/g_{\alpha B}}$
($\alpha \in \{\omega,\rho\}$).  In scenario $\kappa_I$ (consistent
with values derived within QMC) we take these ratios from vector meson
dominance (${\displaystyle \kappa_{\rho N} = f_{\rho N}/g_{\rho N}} =
3.70$). Alternatively (scenario $\kappa_{II}$) 
we take these ratios from the 
Nijmegen potentials (Table VII of Ref.~\cite{Rijken:2010zzb}),
with a larger value of ${\kappa_{\rho N}}$ (= 5.7).
\begin{figure}[!bt]
\centering
\psfrag{f1ylabel}[c][c]{Mass [$M_\odot$]}
\psfrag{f1xlabel}[c][c]{Radius [km]}
\includegraphics[width=0.5\textwidth]{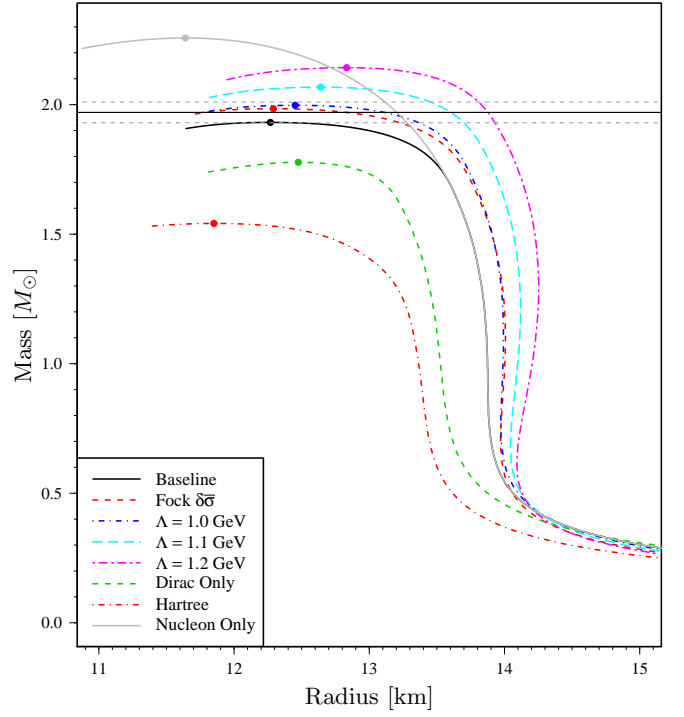} 
\vspace{-7mm}
\caption{Mass vs radius relation for a variety of scenarios described
  in the text. A BPS~\cite{Baym:1971pw} crust has been added in all
  scenarios. We note that the maximum stable neutron star masses
  (indicated by filled circles) for essentially all full Hartree--Fock
  scenarios are compatible with the observations of
  Ref.~\cite{Demorest:2010bx} (denoted by a black horizontal line with
  grey dashed error-band.}
\label{figure:massradius}
\vspace{-5mm}
\end{figure}

Of the baryon-meson coupling constants $g_{\sigma B}(\bar{\sigma}), \,
g_{\omega B}$, and $g_{\rho B}$, only $g_{\sigma B}$ is density
dependent. Its model parameterisation~\cite{Guichon:2008zz} is
dependent on the free nucleon radius, which is taken to be
$R_{N}^{\mathrm{free}}=1.0$~fm -- with an alternate scenario 
having $R_{N}^{\mathrm{free}}=0.8$~fm.
The density dependence is given by
\vspace{-2mm}
\be
g_{\sigma B}(\bar{\sigma}) = -\frac{\partial
  M^{\ast}_{B}}{\partial \bar{\sigma}} \equiv-\frac{\partial
  M^{\ast}_{B}(\bar{\sigma},g_{\sigma N},R^{\mathrm{free}}_{N})}{\partial
  \bar{\sigma}} \ .
  \vspace{-2mm}
\ee

Values of the coupling constants $g_{\alpha N}$ for various mesons
$\alpha$ and a selection of scenarios considered in this work are
presented in Table~\ref{table:couplings}.  The couplings $g_{\omega
  B}$ and $g_{\rho B}$ are expressed in terms of the quark level
couplings
\vspace{-2mm}
\be
g_{\omega B}=n^{B}_{u,d}g^{q}_{\omega} \,\, ; \, \,
g_{\rho B} = g_{\rho N} = g^{q}_{\rho} \, ,
\vspace{-2mm}
\ee
where $n^{B}_{u,d}$ represents the number of light quarks in baryon
$B$.  The $\sigma, \, \omega$ and $\rho$ couplings to the quarks are
constrained to reproduce a saturation energy per baryon of ${\cal
  E}_{\rm sat} = -15.86$~MeV and an asymmetry energy coefficient of
$a_{\rm asym} = 32.5$~MeV at the saturation density
$n_{0}=0.16$~fm$^{-3}$. The $\omega$, $\rho$ and $\pi$ masses are
constrained to their experimental values, whereas the $\sigma$ mass is
taken to be $700$~MeV.

For a compact object in $\beta$-equilibrium we solve the familiar
system of equations for the number densities of the baryons and
leptons~\cite{Glendenning:1997wn}. The lepton energy density and
pressure are given by the usual formulas for a degenerate Fermi
gas. In order to obtain the neutron star properties shown in
Table~\ref{table:couplings}, we solve the
Tollmann--Oppenheinmer--Volkov equations for the gravitational mass
and radius~\cite{Glendenning:1997wn}. The resulting dependence of the
neutron star mass on radius, for a selection of the variations of the
model, is shown in Fig.~\ref{figure:massradius}.
\begin{center}
\begin{table}[tb]
\begin{tabular}{lccccccc}
\hline 
\hline 
\\[-2mm] 
\multirow{2}{*}{Scenario} & \multirow{2}{*}{\ $g_{\sigma N}$\ } &\multirow{2}{*}{\ $g_{\omega N}$\ }&\multirow{2}{*}{\ $g_{\rho}$\ }& $K$ & $R$
 & $M_{\rm max}$ & $n^{\rm max}_{c}$\\
 & & & & (MeV) & (km) & ($M_\odot$) & ($n_0$) \\[1mm]
\hline 
\\[-3mm] 
$\kappa_I$ & 10.42 & 11.02 & 4.55 & 298 & 12.27 & 1.93 & 5.52 \\ 
$\kappa_{II}$ & 10.55 & 11.09 & 3.36 & 299 & 12.19 & 1.93 & 5.62\\ 
$\Lambda = 1.0$ & 10.74 &11.66 &4.68 &305 &12.45 & 2.00 & 5.32\\ 
$\Lambda = 1.1$ & 11.10& 12.33 & 4.84 & 312 &12.64 & 2.07& 5.12\\ 
$\Lambda = 1.2$ & 11.49 &13.06 & 5.03 & 319 & 12.83 & 2.14& 4.92\\ 
$\Lambda = 1.3$ & 11.93 & 13.85 & 5.24 & 329 & 13.02 & 2.23 & 4.74 \\ 
$R = 0.8$ & 11.20 & 12.01 & 4.52 & 300 & 12.41 & 1.98 & 5.38 \\ 
Fock $\delta \bar{\sigma}$ & 10.91 & 11.58 & 4.52 & 285 & 12.29 & 1.98 & 5.5 \\ 
{\footnotesize Dirac Only} & 10.12 &9.25 &7.83 & 294 & 12.56 & 1.79 &5.2\\ 
{\footnotesize Hartree Only} & 10.25 &7.95 & 8.40 & 283 & 11.85 & 1.54 & 6.0\\ 
{\footnotesize Nucleon Only} & 10.42 &11.02 &4.55 &298 &11.64 &2.26 &5.82 \\[1mm] 
\hline 
\hline
\end{tabular}
\vspace{-2mm}
\caption{Meson-nucleon coupling constants determined for our baseline
  scenario `$\kappa_I$' (for which $\Lambda=0.9$~GeV, and $R^{\rm
    free}_{N}=1.0$~fm) and subsequent scenarios in which differences
  from $\kappa_I$ are given in column 1. Also shown are the saturation
  incompressibility, $K$; stellar radius; maximum stellar mass and 
  corresponding central
  density (units $n_0=0.16$~fm$^{-3}$).
  \protect\label{table:couplings}}
\vspace{-5mm}
\end{table}
\end{center}

In Table~\ref{table:couplings} we present the coupling constants,
incompressibilities of symmetric nuclear matter, and stellar
properties, for a number of variations of the QMC model, in each
scenario including the $\sigma,\pi,\omega$ and $\rho$ Fock terms. We note
that in all scenarios $\Xi^-$ hyperons are
present in significant quantities in the the maximum mass stars.
\begin{figure}[!tb]
\psfrag{f2ylabel}[c][c]{\raisebox{3ex}{Pressure [MeV~fm$^{-3}$]}}
\psfrag{f2xlabel}[c][c]{Energy Density [MeV~fm$^{-3}$]}
\includegraphics[width=0.5\textwidth]{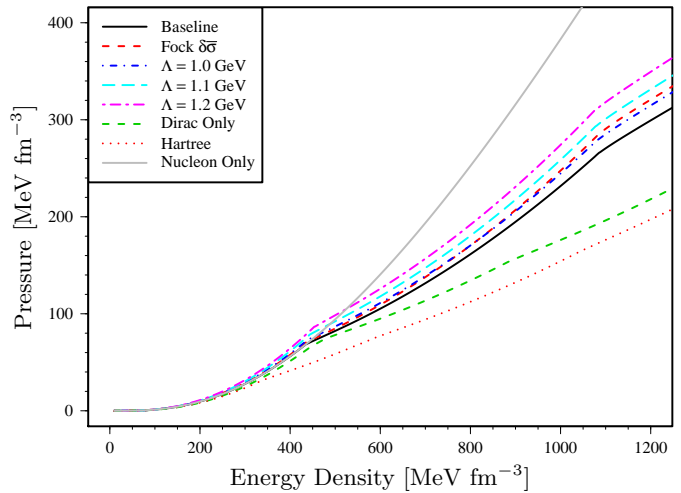} 
 \vspace{-7mm}
 \caption{Equation of state for a variety of
   scenarios. Kinks occur at significant hyperon threshold
   densities. The divergence between the baseline scenario 
($\kappa_I$) and the `Hartree Only' and
   `Dirac Only' scenarios highlights the importance of the $\rho N$
   tensor coupling in Hartree-Fock at high density.}
 \protect\label{figure:eos}
\end{figure}
%

It is remarkable that in all of the scenarios investigated, the
stellar properties are largely consistent, and similar to those
reported by Stone {\it et al.}~\cite{RikovskaStone:2006ta}. Scenarios
in which the maximum stellar mass lies outside of the range
$1.9$--$2.14~M_{\odot}$ correspond to nuclear matter compressibilities
above the upper limit set in the recent
comprehensive analysis of giant monopole resonance
data~\cite{SMS_2012}. While this cannot be true in general, it is
certainly the case for the QMC model.

%
\begin{figure}[!bt]
\psfrag{f3ytoplabel}[c][c]{Species Fractions}
\psfrag{f3ybottomlabel}[c][c]{Fock Energy Density [MeV~fm$^{-3}$]}
\psfrag{f3xlabel}[c][c]{\raisebox{3ex}{Density $n_B$ [fm$^{-3}$]}}
\includegraphics[width=0.5\textwidth]{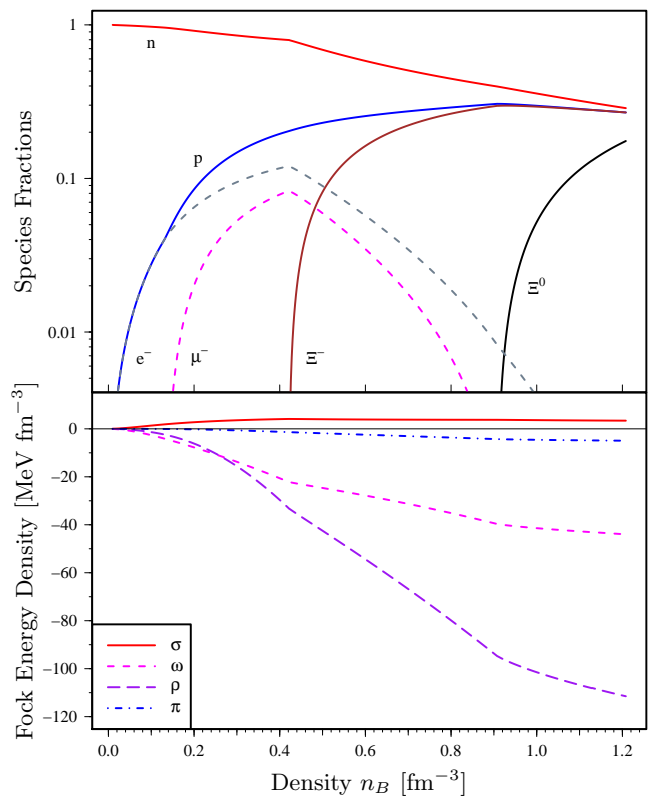}
\vspace{-7mm} 
\caption{Species fractions (upper) and Fock energy densities (lower) for our
  baseline (`$\kappa_I$') scenario.
  \protect\label{figure:eos_params}}
\vspace{-5mm}
\end{figure}
Turning to the effects of the inclusion of the full exchange terms on
stellar properties, we find that the threshold density for $\Xi^{-}$
is lowered, while those of $\Lambda$ and $\Xi^{0}$ are raised, as
demonstrated in Fig.~\ref{figure:eos_params}. In all scenarios there
is a greater splitting between the thresholds of the $\Xi$ 
baryons than that
found by Stone {\it et al.}~\cite{RikovskaStone:2006ta}.  
In our baseline scenario (`$\kappa_I$'), the
$\Xi^{-}$ threshold occurs at $0.42$ fm$^{-3}$, followed by $\Xi^{0}$
at $0.91$~fm$^{-3}$. We find that $\Lambda$ production is not
energetically favoured at densities considered here, in agreement with
Ref.~\cite{Miyatsu:2011bc}. Using the Nijmegen values of tensor
coupling strength (`$\kappa_{II}$'), the $\rho N$ vector coupling is
reduced as the tensor part of the interaction contributes
significantly to the symmetry energy. The EoS is otherwise largely
insensitive to this choice. Similarly, it is insensitive to the
choice of free nucleon radius, despite a moderate impact on the
couplings.

The correction ($\delta \bar{\sigma}$) to the
scalar mean field arising from the Fock terms decreases the
incompressibility by $13$~MeV, yet other observables remain largely
unaltered by this addition.
The cutoff, $\Lambda$, used in form factors (which controls the strength
of the Fock terms) exhibits a more pronounced relationship with the
observables in Table~\ref{table:couplings}.  
Increasing $\Lambda$ beyond 0.9~GeV raises the  
incompressibility with the case 
denoted \mbox{$\Lambda =1.3~{\rm GeV}$} already
exceeding the limit of \mbox{$K < 315$~MeV}. We stress that $\Lambda$
could not take a lower value without impacting the masses of the
$\sigma$, $\omega$, and $\rho$. The $\pi$, however, has a much lower
mass and as such could involve a lower cutoff. We investigated
this possibility but found only a minimal effect on the EoS, as expected
from the small contribution the pion makes to the EoS at high density
(refer to Fig.~\ref{figure:eos_params}). Overall, increases in the
cutoff correlate with increases in both the saturation
incompressibility and maximum stellar mass. 
%
%

We stress that the QMC model does not
predict the appearance of $\Sigma$ hyperons at any density where the
model can be considered realistic.  This is in contrast to a number of
other relativistic models which do predict the $\Sigma$ threshold to
occur, even prior to that of the
$\Lambda$~\cite{SchaffnerBielich:2010am,Weber:2005}.  We note that
Schaffner-Bielich~\cite{SchaffnerBielich:2010am} considered a
phenomenological modification of the $\Sigma$ potential with
additional repulsion, which significantly raised its threshold
density.  In the case of the QMC model the physical explanation of the
absence of $\Sigma$-hyperons is very natural, with the mean scalar
field enhancing the repulsive hyperfine force in the bound
$\Sigma$. Recall that the hyperfine splitting is due to
one-gluon-exchange, which determines the free $\Sigma$--$\Lambda$ mass
splitting in the MIT bag model.

For comparison purposes, we also include a `Nucleon Only' scenario,
in which hyperons are artificially excluded. 
In this case the
EoS is increasingly stiffer at densities above $0.4~{\rm fm}^{-3}$,
leading to a large maximum stellar mass of $2.26~M_{\odot}$,
consistent with many other nucleon-only models.

It is worth remarking that upon inclusion of the tensor coupling, the
proton fraction increases more rapidly as a function of total baryon
density.  This is likely to increase the probability of the direct
URCA cooling process in proto-neutron stars.  As a further
consequence, the maximum electron chemical potential is increased in
this case, which may well influence the production of $\pi^{-}$ and
$\bar{K}$ condensates. Changes to the $\Lambda$ threshold (occurs at
higher density with lower maximum species fraction) reduces the
possibility of H-dibaryon production as constrained by
$\beta$-equilibrium of chemical potentials.

In summary, taking into account the full tensor structure of the
vector-meson-baryon couplings in a Hartree-Fock treatment of the QMC
model gives increased pressure at high density\emdash largely because
of the $\rho N$ tensor coupling\emdash while maintaining reasonable
values of incompressibility at saturation density.  The conceptual
separation between the incompressibility at saturation density and the
amount of pressure or `stiffness' at higher densities is critical. It is
the latter that leads to neutron stars with maximum masses ranging
from $1.90~M_{\odot}$ to $2.14~M_{\odot}$, even when allowance is
made for the appearance of hyperons.  This suggests that hyperons are
very likely to play a vital role as consituents of neutron stars.
%
%
%
%

JRS is pleased to acknowledge the hospitality of the CSSM at the
University of Adelaide, where this work was carried out.  This work
was supported by the University of Adelaide and the Australian
Research Council through grant FL0992247 (AWT) and through the ARC
Centre of Excellence for Particle Physics at the Terascale. JDC was
supported in part by the United States Department of Energy contract
DE-AC05-06OR23177 (under which Jefferson Science Associates, LLC,
operates Jefferson Lab). KT was supported in part by a visiting
professorship at IIP (Brazil).
\vspace{-5mm}
%
%

%
\end{document}